\documentclass{article}
\usepackage{amsmath,amssymb,epsfig}

\newcommand{\tmop}[1]{\operatorname{#1}}
\newcommand{\tmem}[1]{{\em #1\/}}

\begin{document}

\title{Constraining the string scale: from Planck to Weak and back
again}
\author{S. Abel and J. Santiago \\ \\
IPPP, Centre for Particle Theory, Durham  University \\
DH1 3LE, Durham U.K.}
\maketitle


\abstract{
String and field theory ideas have greatly influenced each other since the
so called second string revolution. We review this interrelation paying
particular attention to its phenomenological implications. Our guiding
principle is the radical shift in the way that we think about the 
fundamental scale, in particular the way in which string models 
have been able to accommodate values from the Planck
$M_\mathrm{Pl}\sim 10^{18}$ GeV down to the electroweak scale
$M_{EW}\sim $ TeV.} 

\section{Vacuum degeneracy: vice or virtue? }

There are two purposes to this article. The first is as an overview (for an
experimental audience) of string theory developments in the past 8 
years or so, since
the so-called 2nd string revolution, concentrating on aspects that have to do
with ``phenomenology". Many ideas that are now common, such as large extra
dimensions, have arisen from string theory or at least been inspired by it.
Conversely ideas couched purely in terms of for example extra dimensional
field theory have often guided subsequent string theory developments. For the
lay audience the resulting picture has become
exceedingly obscure, and there is a clear need for some kind of overview. 
This
article is an attempt to meet this need 
by focussing on one particular area where our
ideas have radically changed in the past few years, namely what the
fundamental scale of gravity, or string scale, ought to be. In the first half
of the article therefore, we shall discuss how string models have consistently
been able to accommodate successively lower string scales, until today it is
possible to construct models that have fundamental scales of order a few TeV.

The second purpose of this article is to address the most immediate question
that presents itself to people with a more phenomenological viewpoint, 
namely whether such low scale string models
have anything to do with reality, i.e. do they have any repercussions for
experiment? We shall argue that they indeed do and that flavour changing
effects constrain the string scale to be higher than $10^3$ TeV thereby
eliminating a whole tranche of low scale string models. A more optimistic
phraseology would be to say that experiment is already probing string scales
of order $10^3$ TeV. This indicates, we think, that string theory is finally 
becoming an honest theory, that is one that can be readily disproved by experiment.

A subsidiary purpose of this article, of interest to those concerned with
extra dimensional field theories, is to show how most conceivable ideas that
have been discussed in field theory terms can be constructed and tested in a
stringy set-up. The obvious virtues of the latter are that questions that are
difficult to address (e.g. divergences) or impossible to address (e.g. quantum
effects) in extra dimensional field theory models,  are usually resolved
in string theory.

We begin with a discussion of how we used to estimate the fundamental scale
of quantum-gravity. The familiar estimate is a dimensional one, based on measured
constants of nature
\[ \left. \begin{array}{l}
     G_{} = 6 . 673 \times 10^{- 11} m^3 \tmop{kg}^{- 1} s^{- 2}\\
     h = 1 . 055 \times 10^{- 34} J s\\
     c = 2 . 997 \times 10^8 m s^{- 1}
   \end{array} \right\} \rightarrow L_{\tmop{Pl}} = \sqrt{G h / c^3} = 1 . 61
   \times 10^{- 33} \tmop{cm} \]
The resulting Planck length ($\equiv M_{\tmop{Pl}} = 1 . 22 \times 10^{19}
\tmop{GeV} )$ is the scale at which we used to think quantum-gravity effects
would first make themselves felt.

What can go wrong with this estimate? The crucial point, emphasized in
Ref.{\cite{add}}, is that the energy scale at which we measure $G_N$ is vastly
different from $M_{\tmop{Pl}}$ itself. (This is possible because, alone among
the forces, the effect on gravity of adding extra masses is always positive.)
The implicit assumption is that in between the two scales there are no
abnormally large parameters entering into the physics. In particular for this
discussion, in theories which have extra dimensions that are much larger than
the fundamental scale, the measured Newton's constant can be much weaker than
expected because the gravitational force is diluted by the extra volume.
Indeed the naive relation is
\[ G_4 \sim V^{- 1}_{D - 4} G_D \]
where $V_{D - 4}$ is the volume of whatever extra dimensions our theory
happens to have. (Note that we will for simplicity only consider flat extra
dimensions.) If for example we have a fundamental scale of $M_s \sim 1
\tmop{TeV}$ then $V_{D - 4} \sim 10^{32}$ (in fundamental lengths) gives the
required enhancement factor of $10^{16}$ to the Planck mass. If $D = 10$ then
we would require the extra dimensions to be of order $\tmop{few} \times 10^5
\tmop{TeV}^{- 1} .$

On the other hand gauge forces cannot consistently be allowed to feel the
same extra dimensional volumes. This is because gauge couplings are
dimensionless so that the extra volume would just lead to either
nonperturbatively large or immeasurably small couplings. (They could feel
{\tmem{some}} large volumes however, in which case there is some rescaling
required and the relationships become a little more complicated but similar.)

\begin{figure}[h]
  \begin{center}
    \epsfig{file=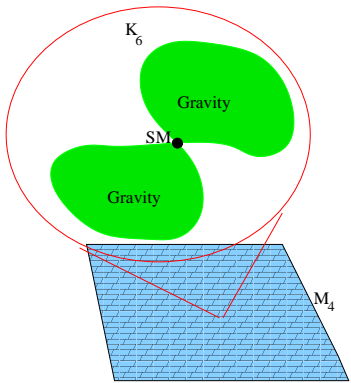,height=5cm}
  \end{center}
  \caption{Brane world picture with 4 large flat dimensions
  represented as a plane and extra small dimensions determining
  the different scales of nature.\label{compactification}}
\end{figure}

The generic picture for significantly changing the scale of quantum gravity is
therefore as shown in Figure~\ref{compactification}. The large flat 4 dimensional space in which we
apparently live is shown as the flat plane. Blowing up any portion of it
reveals an internal space that determines all of the physics (supersymmetry,
particle content and so on).
The fundamental scale can be much lower than the Planck scale if gravity
feels a large internal volume (denoted by green blobs), with the Standard Model 
(SM) fields being confined to some restricted subvolume.

This type of set up is a natural possibility in string theory with its 6
extra dimensions, but {\em large} extra dimensions are a reasonable thing to
consider only because of a feature of string theory that we used to regard as
a problem, namely the {\tmem{vacuum degeneracy problem}}. To summarize, the
problem is that string theory gives no hints as to the shape or size of the
compactified vacua, or even the number of compactified dimensions. So for
example we have no explanation as to why there are 4 large flat dimensions.
More specifically this can be stated as follows. The size and shape of a
particular compactification manifold can be specified by various parameters
(for example the various radii), known collectively as moduli. Choosing a
particular compactification radius corresponds to fixing these parameters.
Since they determine the 4 dimensional physics they should of course be the
same (i.e. Figure~\ref{compactification} should look the same) at every point in $M_4$. However
these parameters correspond to the VEVs of fields in the spectrum that are
left over from the higher dimensional metric. These fields turn out to be
massless, and indeed their potential is completely flat to all orders in
perturbation theory. (In terms of Figure~\ref{compactification}, if for example we perturb the
compactification manifold at a particular point in $M_4$ then all the
neighbouring manifolds are perturbed and so on, and a signal radiates out at
the speed of light in $M_4$; these are the massless particles.) In addition we
are at liberty to set the compactification to be as large as we like, with the
hope that our preferred choice will at some stage be explained by a
non-perturbative contribution to the moduli potential. So when it comes to
lowering the fundamental scale, the vacuum degeneracy problem is seen as a
virtue.

\section{The road to $M_s = M_W$}

We now turn to how this idea has been realized in stringy set-ups. For this we
first need a ``road-map" of string theory in order to orient ourselves; we
begin with the canonical layout of 10 dimensional string theory plus
supergravity shown in Figure~\ref{mtheory}.

\begin{center}
  {\small \begin{figure}[h]
    \begin{center}
      {\scriptsize }\epsfig{file=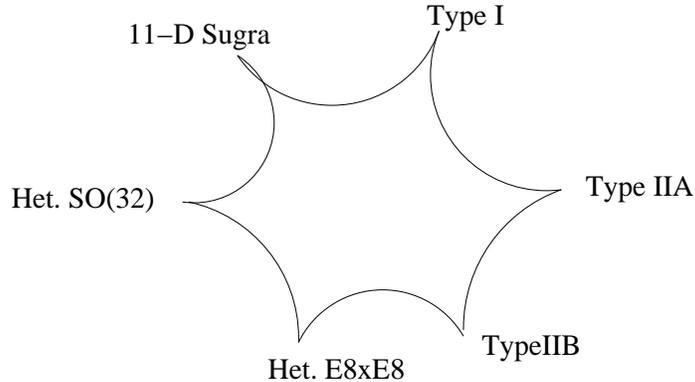,height=5cm}
      \caption{\label{mtheory}M-theory road map}
    \end{center}
  \end{figure}}
\end{center}

Five of the labelled points represent the various perturbative regimes (i.e.
different kinds of string theory) that can be written down in 10 dimensions.
These are Heterotic, and type IIA/B, all of which are theories of closed
strings, and type I which is an $\tmop{SO} ( 32 )$ theory of open strings. In
addition the diagram includes a sixth point representing 11D supergravity. The
triumph of the 2nd string revolution was to demonstrate that by applying
successive duality transformations it is possible to get from any of the 6
perturbative points on this diagram to any other. The conjecture is therefore
that the perturbative theories are simply limits of some nonperturbative
underlying theory which encompasses the whole of this diagram, for which the
search continues. In the meantime one can consider the phenomenological 
possibilities for the 6 theories where we can do perturbation theory.

In the following sections our phenomenological discussion will take us
to all the different corners of this road map. The itinerary is determined by the value
of the string scale in the different models, starting with the most conservative
case of a string scale of the order of the Planck mass in weakly coupled
heterotic models down to GUT string scale (strongly coupled heterotic),
intermediate scale models (type I and II models) and finally discussing the radical
idea of a TeV string scale (in non-supersymmetric models with D-branes intersecting
at non-trivial angles).

\section{
From $M_s \sim M_{\mathrm{Pl}}$ to $M_s \sim M_{\mathrm{GUT}}$:
weakly and strongly coupled heterotic models}

At first sight (i.e. perturbatively) only the Heterotic models seem to be of much use 
for model building. This is because
they alone contain both quantum gravity and gauge fields. Gravity, being a
spin 2 field, requires closed strings which rules out type I. However the type
II models are also ruled out because they {\tmem{only}} contain gravity
multiplets and no gauge fields. Heterotic theories are also closed strings,
but they are a curious combination of supersymmetric and bosonic string
theories. The former can exist in 10 dimensions whereas the latter require 26.
The 16 additional internal degrees of freedom in the bosonic half then become
gauge degrees of freedom in the effective theory; hence the gauge groups must
have rank 16 and indeed anomaly cancellation restricts them still further to
be $E_8 \times E_8$ or $\tmop{SO} ( 32 )$ (the latter turns out to be dual to
the SO(32) of the type I models).

Model building in heterotic strings concentrated on the $E_8 \times E_8$
gauge group. In order to get N=1 supersymmetry in 4 dimensions, the
compactification manifold $K_6$ has to be of a certain type (Calabi-Yau) and
consistency requires a breaking of the gauge group by the compactification.
One attractive route of gauge breaking is then
\begin{eqnarray*}
  E_8 \times E'_8 \longrightarrow \tmop{SU} ( 3 ) \times E_6 \times E_8'
  \longrightarrow & \tmop{MSSM} \times \tmop{hidden} &
\end{eqnarray*}
This route arose from a particularly simple way of satisfying the various consistency
conditions that compactification imposes, which became known as the ``standard embedding''.
The first $E_6$ factor is a potential Grand Unified group whereas the second
$E'_8$ factor forms a hidden sector group. The latter is a potential source of
supersymmetry breaking by for example the condensing of the gaugino of some
hidden sector group at a high mass scale (much like the condensation that
takes place in QCD leading to a $\Lambda_{\tmop{QCD}}$ breaking).

Let us now turn to the question of the fundamental scale. As we have said, in
heterotic models, all degrees of freedom in the perturbative model are the
result of excitations of closed strings. All closed strings can travel
everywhere in the compact space and so both gauge and gravity degrees of
freedom necessarily feel the same compact volume, $V_6$ say.
The Planck scale and the gauge couplings can then be simply computed from
the dimensional reduction of the 10-dimensional theory. In terms
of the string scale, $M_s$, and the heterotic string coupling, $\lambda_H$,
they read
\begin{equation}
M^2_{\mathrm{Pl}}\sim \frac{V_6}{\lambda_H^2} M_s^8,
\label{MPl:vs:Ms}
\end{equation}
and
\begin{equation}
\alpha_{YM} \sim \frac{\lambda_H^2}{V_6 M_s^6}.
\label{gaugecoupling:heterotic}
\end{equation}
These expressions, together with the experimental fact that
$\alpha_{YM}\lesssim 1$, imply, in the case that the heterotic
string remains weakly coupled (\textit{i.e.} $\lambda_H \lesssim 1$),
the following relations between the compactification, string and
Planck scales
\begin{equation}
M_s \sim M_{Pl} \sim V_6^{1/6}.
\end{equation}

\begin{figure}[h]
  \begin{center}
    \epsfig{file=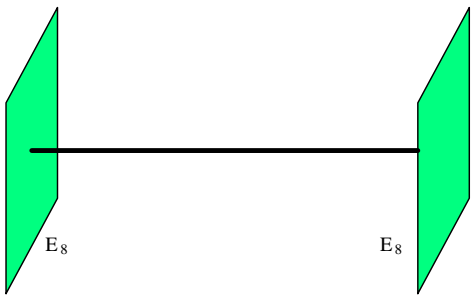,height=4cm}
  \end{center}
  \caption{Horava-Witten construction for the strong coupling limit of
  the $E_8\times E_8$ heterotic string. The green planes represent the
  10-dimensional boundaries of the orbifold $S^1/Z_2$ where each $E_8$
  factor lives, 11-dimensional supergravity  propagates in the bulk.
  \label{HoravaWitten:fig}}
\end{figure}

Things are less simple in the strongly coupled
limit.  The strongly coupled $E_8\times E_8$ heterotic is only
tractable thanks to the fact that, as Horava and Witten
showed~\cite{Horava:Witten}, it is described by 11-dimensional
supergravity compactified on an $S^1/Z_2$ orbifold. Based on
anomaly cancellation arguments they argued that an $E_8$ gauge
group lives on each of the two 10-dimensional orbifold fixed planes
whereas gravity lives in the 11-dimensional bulk as sketched in
Fig.~\ref{HoravaWitten:fig}.
In the case of strong coupling, the radius of the orbifold $R_{11}$
is larger than the compactification scale of the 6 extra dimensions.
It is therefore possible to consider the compactification of this
theory down to 4 dimensions in two steps, with an intermediate
5-dimensional model compactified on an orbifold.

The 11-dimensional action takes the form
\begin{equation}
S=\frac{1}{2\kappa_{11}^2} \int \mathrm{d}^{11}x\, \sqrt{g} R
-\sum_{i}\frac{3^{1/3}}{4\pi(2\pi \kappa_{11}^2)^{2/3}}
\int \mathrm{d}^{10}x\, \sqrt{g} \mathrm{Tr}F_i^2+\ldots,
\end{equation}
where $\kappa_{11}$ is the 11-dimensional gravitational constant
and $i$ runs over the two 10-dimensional fixed planes where the two 
$E_8$ groups live. Compactifying down to five dimensions
(with a compact volume $V_6$) and then to four dimensions we
can write the fundamental 11-dimensional constant, $M_{11}=
2\pi(4\pi \kappa_{11}^2)^{-1/9}$ and the radius of the
11-th dimension, $R_{11}$, in terms of 4-dimensional quantities,
\begin{equation}
M_{11}=(2\alpha_{GUT} V_6)^{-1/6}, \quad
R_{11}^2=\left(\frac{\alpha_{GUT}}{2}\right)^{3}V_6 M_{\mathrm{Pl}}^4.
\end{equation}
It is now possible to have
\begin{equation}
M_{11} \sim V_6^{-1/6} \sim M_{\mathrm{GUT}}\sim 10^{16} \mbox{ GeV},
\end{equation}
and therefore $R_{11}^{-1} \sim 10^{13}-10^{15}$ GeV.

Thus we have seen how the heterotic string can accommodate a fundamental
scale  of the order of the Planck mass in the weak coupling limit, 
or GUT scale in the strong coupling limit. 
In the following sections we shall see how
the existence of D-branes in type I and II theories allows an even greater
reduction in the fundamental scale.

\section{Intermediate models}

\begin{figure}
\begin{center}\includegraphics[%
  scale=0.4]{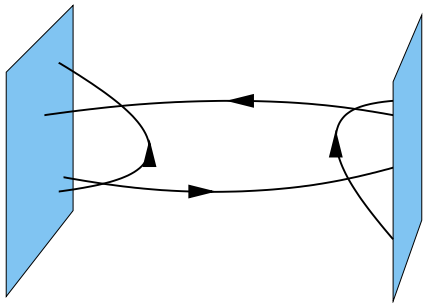}\end{center}

\caption{\label{cap:Gauge-bosons-of}D-brane realization of a
U(2) gauge group.}
\end{figure}

The arrival of the large extra dimension idea stimulated interest
in the other variants of string theory as model building tools. In
particular attention turned to the type I and type II theories which
have in their nonperturbative spectrum objects known as Dirichlet
branes \cite{d-branes,Gimon:1996rq}. 
These can be built like monopoles from the effective field
theory, and are membrane-like and fully dynamical, with a typical
surface tension and a width of order the fundamental scale. They have
$p$ dimensions on their world volume where $p=1,3,5,7,9$ for type
IIB, $1,5,9$ for type I and $0,2,4,6,8$ for type IIA. The interesting
feature of D-branes from a model builder's point of view is that open
strings can end on them and this can generate gauge groups in the
following way. Associated with an open string end point is an index,
the Chan-Paton index. If there are a few branes together, the index
simply labels the branes to which the open string is attached. If
we consider two branes for example, the endpoints can be attached
in one of 4 ways as in Figure \ref{cap:Gauge-bosons-of}.

What do we see when we observe this from 4 dimensions? Remember that
from the 4 dimensional point of view we need to arrange things such
that the compactified space is the same everywhere. In particular
the brane must be lying in the large $M_{4}$ space that we observe
in order for the open string to be able to travel along it (otherwise
it would be stuck at a single point in $M_{4}$. So the branes must
have $p\geq3$. (If $p=3$ the branes appear as points in the compactified
space.) Given this, the open strings may freely propagate in $M_{4}$
but have 4 internal degrees of freedom corresponding to the adjoint
of U(2). It also turns out that the strings have to have an excitation
from the brane volume giving them a Lorentz (gauge boson) or internal
(matter field) index. Finally a remarkable feature of D-branes is
that they break only half the supersymmetry. Thus the original theory
which has $N=8$ supersymmetry in 4 dimensions (if the compactified
space is toroidal) ends up being $N=4$. We thus end up with an $N=4$
theory with $U(2)$ gauge group. In order to reach a more phenomenogically
interesting $N=1$ configuration, the compactified space $K_{6}$
can be chosen in such a way that the supersymmetry is already partially
broken before the D-branes are added. A type of compactification which
is particularly easy to work with are orbifolds - spaces with curvature
singularities at fixed points of the orbifolding (like the corners
on cushions).

Before we start throwing branes together at random, we need to take
care of some consistency conditions. The most important of these for
D-branes are the Ramond-Ramond tadpole conditions. Every D-brane has
a ``Ramond-Ramond'' (RR) charge, and couples to Ramond-Ramond
fields that exist in the closed string spectrum (that is they are
closed string excitations that are present in the type I or type II
theory even before the D-branes are added in). Since these are closed
string states they do not care about the presence or otherwise of
the D-branes. In a toroidal compactification they propagate throughout
the entire compactified volume. Curvature singularities, for example
when the compactified space is an orbifold, introduce a second type
of {}``twisted'' RR field that is confined to the fixed points. The
RR fields behave rather like gravitons and dilatons and form part of
the gravitational spectrum. However they differ in the respect that
flux lines of Ramond-Ramond fields must be absorbed in a compact space
otherwise the theory is rendered entirely inconsistent. One has to
be careful therefore to choose the arrangements of D-branes such that
the flux lines are all absorbed. Once this requirement is satisfied,
other requirements such as anomaly cancellation are usually satisfied
as well.

\begin{figure}
\begin{center}\includegraphics[%
  scale=0.5]{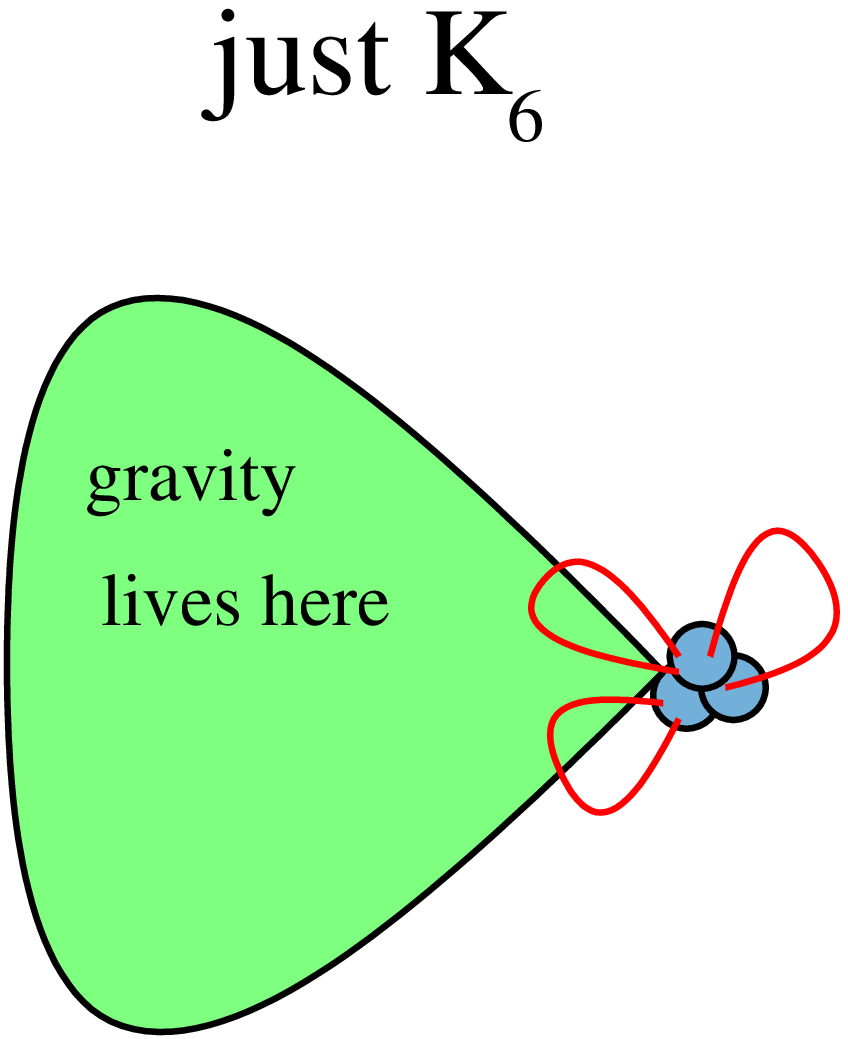}\end{center}
\caption{\label{cap:Schematic-picture-of}Schematic picture of the bottom-up
approach. The small blue points represent the local configuration of D-branes
leading to the MSSM
whereas the large green blob represent the global structure, less important
from a phenomenological viewpoint.}
\end{figure}

These requirements led to an approach to model building which became
known as {}``bottom-up'' \cite{bottom-up}. Consider what are the
important features of any model from the point of view of phenomenology.
The leading factors are those things that have to do with the gauge
groups, particle content, number of generations and so on. Secondary
factors are things that have to do with supersymmetry breaking, the
cosmological constant etc. The latter are things whose eventual properties
are intertwined with gravity. As such their influence on phenomenology
is less important. In a large extra dimension set-up, the correspondence
with the configuration in the compactified space is rather direct.
The primary factors have to do with the local arrangements of D-branes
around, for example, some orbifold fixed point, whereas the secondary
factors are all associated with objects far away in the bulk of the
compactified space. For example a {}``hidden'' sector can be included
consisting of a collection of branes at some \textit{other} fixed
point far away in the compactified space. The communication to the
visible sector then has to be through the bulk, and will get the same
volume suppression as that felt by gravity. This is shown schematically
in Figure \ref{cap:Schematic-picture-of}. The points represent for
example D3 branes localized at some point in the compactified space
with twisted RR flux cancelled locally. These are chosen in such a
way that the visible sector is the MSSM. Gravity and the untwisted
RR fields live in the bulk of the compactified space. These details
and in particular the details of untwisted RR flux cancellation are
less well determined.

\begin{figure}
\begin{center}\includegraphics[%
  scale=0.5]{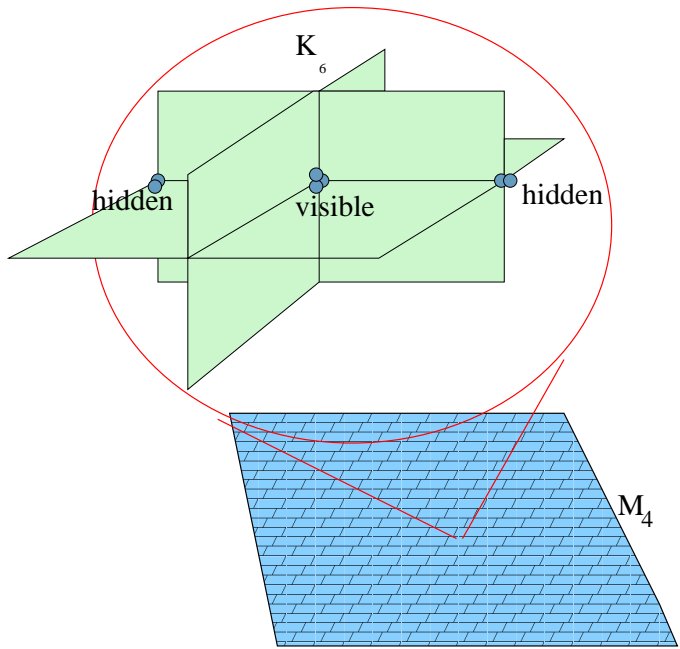}\end{center}
\caption{\label{cap:Set-up-for-the}Set-up for the bottom up approach. The
visible sector consists of 3-branes at a fixed point in $K_{6}$.
D7 branes have to be included passing through this fixed point to
cancel local RR-tadpoles. Global absence of tadpoles requires additional
branes and/or anti-branes in the bulk, or possibly something else
entirely. }
\end{figure}

The bottom-up approach begins therefore by focussing on the local
MSSM configuration. We assume an intermediate fundamental scale of
\begin{equation}
M_{I}\sim\sqrt{M_{W}M_{Pl}}\sim10^{11}\, GeV.\end{equation}
This scale is familiar from the hidden sector supersymmetry breaking
communicated by gravity and had been suggested earlier on more general
grounds to do with supersymmetry breaking and mediation by gravity.
First a set of D-branes is included at some fixed point of $K_{6}$
with all the necessary elements to make up the standard model gauge
group and leave $N=1$ supersymmetry in the visible sector. This can
for example be a set of D3-branes lying on top of each other at a
single point in $K_{6}$, but with their world volumes filling the
whole of $M_{4}$ (as of course we require if the open strings on their
world volumes are able to travel anywhere in $M_{4}$). We then need
to satisfy the requirements of local RR-tadpole cancellation. That
is we need to add in additional branes (D7 branes for example) such
that the {}``twisted'' RR-tadpoles cancel but locally supersymmetry is preserved. 
This puts a constraint on the angles at which the branes can interesect
(for example that the D7 branes intersect at right angles). 
This arrangement takes
care of the local consistency conditions, however one should also
take care of the global RR-tadpoles and make sure those fluxes cancel
as well. This however can be done by adding other D-branes and anti-D
branes elsewhere in the bulk or may be done in some other way. From
the point of view of 4D phenomenology therefore, the particular way
in which the global tadpoles are cancelled affects only the hidden
sector, and consequently the soft supersymmetry breaking and cosmological
constant. A consistent set-up is shown schematically in Figure \ref{cap:Set-up-for-the}.
This figure shows the global RR flux being absorbed by anti-branes,
but the set-up can be entirely different away from the visible sector
without affecting the MSSM set-up directly.

The reason for the particular choice of the intermediate scale can
now be made clear. The additional ingredients required to ensure global
tadpole cancellation generally break supersymmetry. Since it is only
the global configuration that breaks supersymmetry, the net effect
is the same as hidden sector supersymmetry breaking communicated by
gravity and we must choose the fundamental scale accordingly. In other
words, the volume of the bulk can be responsible for the large Planck
scale and the dilution of supersymmetry breaking effects only if $M_{s}\sim M_{I}$.
The precise dependences on volumes can be derived from the reduction
of the effective 10 dimensional type I action to 4 dimensions~\cite{Ibanez:1998rf}.
We begin with the Planck mass relation to the total compact volume
\begin{equation}
V_{K_{6}}=\lambda_{I}^{2}\frac{M_{P}^{2}}{M_{s}^{2}},\label{eq:volk6}\end{equation}
where $\lambda_{I}$ is the string coupling. To get an idea of what
this has to be, we can look at the effective gauge coupling $\alpha_{p}$
on a $p$-dimensional brane. The gauge interactions are proportional
to the string coupling but are diluted by the volume of the branes
in the compactified space, $V_{p-3}$, since the gauge bosons are
free to roam anywhere in this volume. Hence \begin{equation}
\alpha_{p}\sim\frac{\lambda_{I}}{V_{p-3}}.\label{eq:ap}\end{equation}
Substituting Eq. (\ref{eq:ap}) into Eq. (\ref{eq:volk6}) gives us \begin{equation}
\alpha_{p}M_{P}^{2}=M_{s}^{2}\frac{V_{9-p}}{V_{p-3}},\end{equation}
where $V_{9-p}$ is the co-volume (i.e. the volume orthogonal to the
$p$ brane). Any process we care to calculate that breaks supersymmetry,
such as a contribution to the scalar mass-squareds communicated via
closed string modes from an anti-brane, feels the same volume dependence\begin{equation}
m_{SUSY}^{2}\sim M_{s}^{2}\frac{V_{p-3}}{V_{9-p}}.\label{eq:msusy}\end{equation}
The dilution due to the co-volume $V_{9-p}$ is obvious. The $V_{p-3}$
enhancement factor arises from the sum over Kaluza-Klein (momentum)
modes in the brane volume and is essentially the same factor as 
arising that arising in $1/\alpha_p$.  
Essentially this is like a phase space
factor. (As a rule-of-thumb, one can use the fact that if we invert
a radius, $R_{i}\rightarrow1/R_{i}$, we also turn that dimension
from a brane dimension into a dimension orthogonal to the brane or
vice-versa, and also change the dimensionality of the brane, $p\rightarrow p\pm1$.
Hence the volumes must appear as the ratio of brane volume to co-volume,
$V_{p-3}/V_{9-p}$.) There is no $1/\lambda_I $ contribution 
as there is in the tree level Yang-Mills terms (hence the equation for 
$\alpha_p$) because the diagrams that contribute to $M_{SUSY}$
are one-loop and $\lambda_I$ acts like a loop expansion parameter. 

Now, for reasonable phenomenology we would like
$M_{SUSY}\sim M_{W}$ so that from the above, and assuming that we
have $\alpha_{p}\sim1$ we need \[
M_{s}^{2}\sim M_{W}M_{P}\]
as expected, and consequently a volume ratio \begin{equation}
\frac{V_{p-3}}{V_{9-p}}\sim\frac{M_{W}}{M_{P}}\label{eq:vddvnn}\end{equation}
The beauty of the bottom-up approach is that is allows us to disregard
those parts of the construction that are not vital to phenomenology.
For example there is a question of global validity of these models
due to the fact that there are uncancelled tadpoles of another kind,
namely NS-NS tadpoles. These however can be absorbed dynamically by
adjusting the background (i.e. $K_{6}$) and their presence does not
automatically render the theory inconsistent \cite{fischlersusskind}.
Although this effect may make the theory intractible on a global scale,
it may still be a reasonable approximation to assume a nice (tractable)
flat or orbifold background near the visible sector branes, where
we can still calculate, for example, interactions.

\begin{figure}
\begin{center}\includegraphics[%
  scale=0.5]{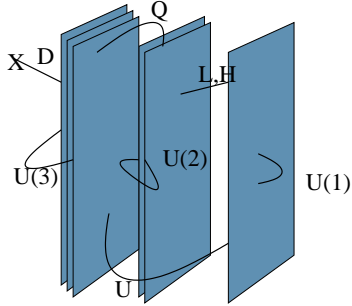}\end{center}
\caption{\label{cap:Local-arrangement-of}Local arrangement of states on D3-branes
leading to the MSSM. }
\end{figure}
Let us turn briefly to the local arrangement of branes that yields
the visible sector particle content and gauge group. This is often
represented as in Figure \ref{cap:Local-arrangement-of}. The Figure
shows the arrangement of $D3$ branes at a particular fixed point
in $K_{6}$. The branes are extended in $M_{4}$ and fixed in $K_{6}$
so that two of the dimensions shown are in $M_{4}$ and the dimension
orthogonal to the branes should be in $K_{6}$. In addition the branes
are on top of each other. (Any separation of branes translates into
a mass for the relevant states due to the stretching energy.) There
are three stacks of branes corresponding to a gauge group $U(3)\times U(2)\times U(1)$.
The gauge states are those strings with ends attached on a single
stack of branes. The matter states correspond to strings stretched
between different stacks of branes and consequently appear (in this
simple example) in the bifundamental. Thus we can identify strings
stretched between the $U(3)$ and $U(2)$ stacks with left handed
quarks, $Q_{L}$, between the $U(2)$ and $U(1)$ branes with left
handed leptons and higgses, and between the $U(3)$ and $U(1)$ branes
with right handed quarks. The gauge groups contain too many $U(1)$
factors, and the final reduction down to a single $U(1)_{Y}$ of hypercharge
comes about because there is only one linear combination of $U(1)'s$
that is anomaly free. Of course string theory is a consistent theory,
and there should be no anomalies at all. But the way in which string
theory cancels the anomalies makes the \textit{naively} anomalous
$U(1)'s$ massive, and one expects that the anomalous combinations
will be broken. Remarkably the states turn out to have the hypercharge
assignments of the SM.

The bottom up approach has a number of advantages, many of which were
outlined in Refs.\cite{bottom-up,allanach}. For example the prediction
of an intermediate fundamental scale is interesting for a number of
reasons. It is a natural realization of hidden sector supersymmetry
breaking communicated by gravity. The model provides axions with just
the right Peccei-Quinn scale to allow an axion solution to the string
CP problem. In addition the see-saw mechanism for neutrino masses
is consistent with a fundamental intermediate scale, and so on.
One of the disadvantages of the bottom-up approach is that, by its
very nature it is difficult to make concrete predictions of phenomenological
implications. This is because the approach begins with a visible sector
that resembles the MSSM and, by construction, aspects such as supersymmetry
breaking have to do with the global configuration over which we assume
very little control. 

What then can be said about the emergent phenomenology?
In the next subsections we will pick out a couple of areas that are
currently exercising us, where the bottom-up approach can make generic
predictions. The first concerns a little considered possibility in
models that have several $U(1)$ factors, namely millicharged particles.
The second related area has to do with the generic properties of supersymmetry
breaking. At the end of this section we shall summarize where other
progress has been made on this question.

\subsection{Kinetic Mixing and Millicharged particles}

\begin{figure}
\begin{center}\includegraphics{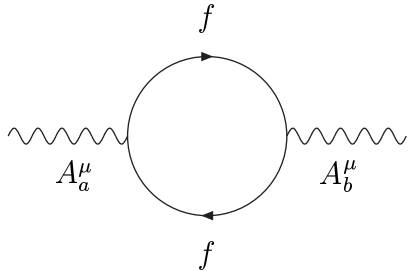}\end{center}
\caption{\label{cap:Contributions-to-Kinetic}Contributions to Kinetic Mixing
in field theory.}
\end{figure}

Millicharged particles are a possibility in any theory that has a number
of $U(1)'s$, as string theories with stacks of D-branes generally
do. The phenomenon that gives rise to this effect is known as Kinetic
Mixing. Consider for example a field theory that has, in addition
to some visible $U(1)_{a}$, a $U(1)_{b}$ factor in the hidden sector.
Kinetic Mixing happens when the hidden $U(1)_{b}$ couples to the
visible $U(1)_{a}$ through the diagram in Figure \ref{cap:Contributions-to-Kinetic}.
The fields in the loop correspond to heavy states that do not appear
in the low energy theory.

\begin{figure}
\begin{center}\includegraphics{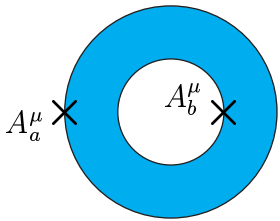}\end{center}
\caption{\label{cap:String-Contributions-to-Kinetic}Contributions to Kinetic
Mixing in string theory}
\end{figure}
This diagram, proportional to $Tr(Q_{a}Q_{b})$ , results in a Lagrangian
of the form \cite{Holdom:1986ag}\begin{equation}
\mathcal{L}_{YM}=-\frac{1}{4}F_{a}^{\mu\nu}F_{a\mu\nu}-\frac{1}{4}F_{b}^{\mu\nu}F_{b\mu\nu}+
\frac{\chi}{2}F_{a}^{\mu\nu}F_{b\mu\nu}\end{equation}
 The most immediate consequences of this type of mixing were first
studied by Holdom \cite{Holdom:1986ag}. On diagonalizing the Yang-Mills
lagrangian, one finds that the hidden sector fields charged only under
$U(1)_{b}$ pick up a small charge of order $\chi$ under the visible
sector $U(1)$. The bounds on such particles are very severe, especially
if they are massless. In fact if $m_{hidden}<m_{e}$ there is a constraint
of $\chi<10^{-15}$ coming from astrophysical bounds (specifically
plasmon decay in Red Giants). Direct but much weaker experimental
bounds of $\chi<10^{-4}$ are found from orthopositronium decay as
well as a number of other accelerator and astrophysical sources \cite{Davidson:2000hf}.

This phenomenon generally arises in the intermediate models \cite{ben}
because they include a hidden sector of anti D-branes in the bulk.
Anti-branes and branes couple by exchanging closed string modes through
the bulk. However a closed string exchange (which resembles a cylinder)
can also be interpreted as an open string stretched between brane
and anti-brane going in a loop as shown in Figure \ref{cap:String-Contributions-to-Kinetic}.
The modes in the loop are heavy because they have a stretching energy
proportional to their length (i.e. the distance between brane and
anti-brane since they are stretched between them). Importantly the
presence of anti-branes breaks supersymmetry. Therefore the one loop
diagrams do not cancel (as they would in a supersymmetric configuration
with just parallel branes for example), and there is a residual contribution
to kinetic mixing and hence $\chi$.

Going back to the closed string exchange picture, it is (almost) obvious
that the diagrams receive the same sort of volume suppression as the
gravitational diagrams that lead to a large $M_{P}$. That is, the
coupling $\chi$ is diluted by co-volume $V_{9-p}$ (i.e. the volume
orthogonal to the D-branes) and enhanced by the brane volume $V_{p-3}$.
However the normal Yang-Mills couplings are enhanced by the same $V_{p-3}$
factor since the gauge bosons travel the entire volume of the brane
(indeed this is where Eq. (\ref{eq:ap}) comes from). The suppression
of the Kinetic Mixing term relative to the normal Yang Mills terms
is therefore suppressed only by the co-volume but has an extra $\lambda_I$ 
factor because it is one-loop. Using the expression 
for $\alpha_p$ we can therefore write
\begin{equation}
\chi\sim Tr(Q_{a}Q_{b})\frac{\alpha_p V_{p-3}}{V_{9-p}}.\end{equation}
Comparing this to Eq. (\ref{eq:vddvnn}) we find \[
\chi\sim \alpha_p Tr(Q_{a}Q_{b})\frac{M_{W}}{M_{P}}.\]
Thus independently of $p$ this crude estimate gives 
the expected value of $\chi$ to be just
below the bounds extracted from astrophysical considerations. 
However there are a number
of factors that we have glossed over here for simplicity, and the
situation can be slightly more complicated due to uncancelled NS-NS
tadpoles, extremely non-degenerate extra dimensions and so on. One of the 
most important factors is that the volume dependence of $\chi $ does 
not quite go as the co-volume. Indeed if the radius of the co-volume is 
$R$ then the supersymmetry breaking mass-squared depend on 
$1/V_{9-p}\sim R^{p-9}$ 
whereas $\chi \sim R^{p-7}$ which merely 
reflects the fact that the potential is just that of gravitational attraction
(e.g. in D space-time dimensions instead of 10 we have $R^{p-D-3}$ and get 
the familiar $1/R$ if we have particles/D0-branes in $D=4$.).
The end result will be a more complicated dependence on $p$ and an enhanced 
millicharged particle effect with some cases being ruled out. For 
further details the reader is directed to Ref. \cite{ben}.

\subsection{Kinetic mixing and SUSY breaking }

\begin{figure}
\begin{center}\includegraphics[%
  scale=0.6]{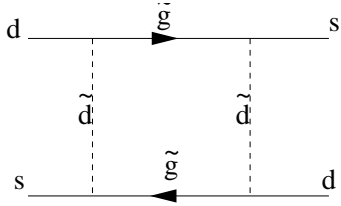}\end{center}
\caption{\label{cap:Supersymmetry-breaking-contribution}Supersymmetry breaking
contribution to $\Delta M_{K}$}
\end{figure}

\begin{figure}
\begin{center}\includegraphics[%
  scale=0.6]{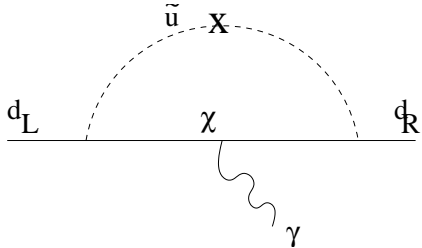}\end{center}
\caption{\label{cap:Supersymmetry-breaking-contribution-to-edm}Supersymmetry
breaking contribution to $d-$EDM}
\end{figure}

The consequences of $\chi\sim M_{W}/M_{P}$ are extremely interesting
for supersymmetry breaking and it is to this that we now turn. Before
we consider the specifics of kinetic mixing in detail we should mention
some of the problems of supersymmetry breaking in generic models that
we would like to be able to solve. One of the most enduring problems
arises from possible flavour non-universality in the supersymmetry
breaking terms. This leads to large violations of flavour from diagrams
such as those in Figure \ref{cap:Supersymmetry-breaking-contribution}
and large EDM contributions such as those in Figure \ref{cap:Supersymmetry-breaking-contribution-to-edm}.

 In the former example, the flavour changing is driven by generational
mass differences, $m_{\tilde{s}}^{2}-m_{\tilde{d}}^{2}$. In the second
example, even though it is CP violating and flavour conserving, large
contributions come from non-universal $A$-terms in the lagrangian
once the fermions are rotated to their mass basis. These problems
are known collectively as the supersymmetric flavour and CP problems.
In the bottom-up configuration it is very difficult to say anything
general about them. However one can identify new possibilities for
solving them.

This is where Kinetic Mixing can play and important role. Dienes \emph{et
al} pointed out that Kinetic Mixing can contribute significantly and
even dominantly to supersymmetry-breaking mediation \cite{Dienes:1996zr}.
This results in additional contributions to the scalar mass-squared
terms proportional to their hypercharge (assuming the visible sector
group is $U(1)_{Y}$, otherwise whatever the charge under $U(1)_{a}$
is) as follows. The supersymmetry breaking in the hidden sector is
assumed to be maximal, and one expects a non-zero VEV for the D-terms
of the hidden $U(1)_{b}$ of order \begin{equation}
\langle D_{b}\rangle\sim M_{s}^{2}.\end{equation}
Upon kinetic mixing, the fields in the visible sector see this VEV
through their $U(1)_{b}$ millicharges. That is the scalars pick up
effective mass-squareds order\begin{equation}
\tilde{m}^{2}\sim Q_{a}\chi M_{s}^{2}.\label{eq:mchi}\end{equation}
 The interesting feature of the kinetic mixing in intermediate models
is that it gives\[
\tilde{m}^{2}\sim Q_{a}\chi M_{I}^{2}\sim Q_{a}M_{W}^{2}.\]
The volume factors required to produce the large Planck mass therefore
match those required to suppress $\chi$ to be $M_{W}/M_{P}$. Furthermore
the mass-squared terms are proportional to hypercharge and is therefore
generation independent. What we have found in intermediate models
therefore is a source of degenerate mass-squareds of order $M_{W}^{2}$
that can be used to solve or at least ameliorate some of the problems
of generic supersymmetric models. As pointed out by in ref\cite{Dienes:1996zr},
Kinetic Mixing to hypercharge cannot be the only source of mediation,
as some of the mass-squareds would have to be negative however the
problem should be reduced. Alternatively one could invoke a second
$U(1)_{a}$ group in the visible sector that provides additional positive
mass-squared contributions.

It is interesting to contrast the Kinetic Mixing here with that in
Ref.\cite{Dienes:1996zr}. In a model with a string scale of say $M_{s}\sim M_{GUT}$,
including $\chi\sim M_{W}/M_{P}$ looks like a very unnatural fine
tuning according to the criterion of t'Hooft. These authors therefore
focused on placing an upper limit on $\chi$ in order to avoid destabilizing
the gauge hierarchy (\emph{i.e.} to avoid supersymmetry breaking in
the visible sector much larger than 1 TeV). The appropriate limit
on $\chi$ then depended on the scale of supersymmetry breaking in
the hidden sector which in turn depends on the other sources of mediation
(\emph{e.g.} gravity or gauge). The conclusion was that generic models
with gravity mediation would have disastrously large Kinetic Mixing
if the hidden sector contained additional $U(1)$'s. The relevant
bound to avoid destabilizing the hierarchy is of course $\chi<M_{W}/M_{P}\sim10^{-16}$
as is clear from Eq. (\ref{eq:mchi}). Clearly values of $\chi$ much
larger than this will produce scalar masses much greater than $M_{W}$.
In heterotic strings the situation is ameliorated somewhat because
the gauge groups are usually unified into some non-abelian GUT groups.
The Kinetic Mixing only arises due to mass splittings once the GUT
groups are broken, and one finds typical values of $\chi\sim10^{-9}$;
much less than 1 but still large enough to destabilize the hierarchy.

\subsection{Other aspects}

There are a number of other areas where changing the fundamental scale
has a significant effect. In particular the structure of supersymmetry
breaking is quite different due to the different scales involved in
renormalization group running. The most obvious impact is on the spectrum
of the supersymmetric scalars discussed in Ref.\cite{intspectrum+dm}.
These works also focussed on the affect on the neutralino dark matter
cross section, and this was picked up on and extended in Refs.\cite{intdm,intrare+dm}
where it was emphasized that there are regions of parameter space
where the neutralino-nucleon cross section is significantly enhanced,
making dark matter detectable in current experiments. The second reference
also extends the discussion to rare processes, and finds that there
is a significantly different correlation between dark matter and rare
processes. In particular lowering the string scale changes $B$ decay
rates, in particular $B_{s}\rightarrow\mu^{+}\mu^{-}$.

One aspect about which intermediate models have something interesting
to say is an alternative solution to the flavour and CP problems in
supersymmetry, the so-called dilaton domination solution. Dilaton
domination is a pattern of supersymmetry breaking that arises when
the main contribution to supersymmetry breaking is the dilaton field.
Since the dilaton is part of the gravity multiplet it couples universally
to all matter and in particular to all the generations. Consequently
supersymmetry breaking driven by the dilaton solves the supersymmetry
flavour and CP problems since the soft supersymmetry breaking that
apears in the effective four dimensional theory is necessarily flavour
universal. However in conventional MSSM models with unification at
the GUT scale, dilaton domination is excluded because of cosmological
considerations. Specifically the electro-weak vacuum is unstable to
decay into deeper minima that break charge and/or colour 
\cite{ccb,allanach,intccb}. (See also\cite{abelallanach} for a
discussion on scales.)
It is possible to live with such an instability if the decay time
is long enough (i.e. longer than the age of the Universe) but it is
more usual for the existence of a charge and colour breaking (CCB)
minimum to be taken as grounds for a model to be excluded. CCB minima
are driven by the negative mass-squared of the higgs fields (which
are also responsible for the successful prediction of electroweak
symmetry breaking of course) which in turn is driven by the effect
of the large top-quark Yukawa in the renormalization group running.
Because of the important effect of the renormalization group, things
are rather different in intermediate scale models essentially because
of the shorter interval (in energy scales) that the soft-supersymmetry
breaking parameteres have to run. This is discussed in Ref.
\cite{allanach,intccb}.

Finally we should mention the impact of lowering the string scale
on the so-called fine tuning parameters of supersymmetry. This was
discussed in Ref.\cite{intfinetuning} which concluded that lowering
the string scale actually increases the amount of fine-tuning required
to produce the correct $M_{Z}$ whilst having relatively heavy scalar
masses.

\section{$M_s \sim$TeV: Branes at angles }

Following the progression down to lower string scale models,
we will discuss in this section a class of models that represent,
within a bottom-up approach, realistic string models with many of the
features of the SM, allowing in principle for a very low string scale.
Our main aim in this review is to account for their phenomenological
features, their realistic structure and, especially, their flavour structure,
which, as it turns out, provides the deepest probe of this kind of
models and the most stringent constraints on the string scale as well.

Models with D-branes intersecting at non-trivial angles~\cite{Berkooz:1996km}
(see~\cite{Bachas:1995ik} for an earlier application of the same
idea, in the dual version of branes with fluxes, to supersymmetry breaking),
have a number of very appealing phenomenological features such as
for instance four-dimensional chirality or a reduced amount of symmetries
(both gauge and supersymmetries) among many others. 
One particularly important feature that these models have 
is an attractive explanation for family replication. Specifically the matter 
fields
correspond to the string states 
at the intersections that are stretched between two branes. There are then 
three generations simply because the branes are wrapped so that 
each type of 
intersection appears three times, with a repeated set of multiplets 
stretched between the branes at the intersections. 

Configurations with branes at angles typically break all the supersymmetries
(supersymmetric configurations have been constructed~\cite{Cvetic:2001tj}
but they are very constrained and minimal models are very difficult to obtain)
and therefore a very low string scale $\sim$ TeV is required.
The first semi-realistic models were constructed 
in~\cite{Blumenhagen:2000wh} and soon
after in \cite{Forste:2001gb} and~\cite{Aldazabal:2000dg}
(see~\cite{Blumenhagen:2000fp} for some related technical developments).
These initial models presented additional gauge symmetries or matter
content beyond the ones in the SM. The first models containing just
the SM were presented in~\cite{Ibanez:2001nd}.
Since then, a great deal of effort has gone into into the study of the
consistency and stability~\cite{Rabadan:2001mt} and phenomenological
implications of intersecting brane models, from the construction
of supersymmetric models~\cite{Cvetic:2001tj}, gauge symmetry breaking
~\cite{Cremades:2002cs}, GUT or realistic SM constructions~\cite{Kokorelis:2002ip,Ellis:2002ci}
to cosmological implications~\cite{Garcia-Bellido:2001ky}.
In the following we will review some of these developments
paying particular attention to their flavour structure~\cite{Cremades:2003qj,
Abel:2003fk,Chamoun:2003pf} and its profound experimental implications.

\begin{figure}[h]
\begin{center}
\epsfig{file=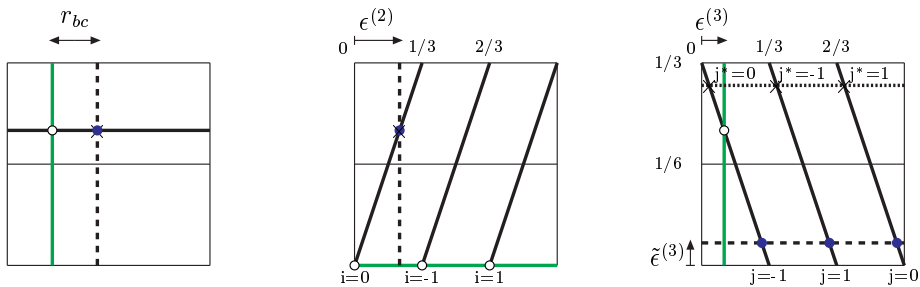,width=\textwidth}
\end{center}
\caption{Brane configuration in the model discussed in the text.
The leptonic sector is not represented while the baryonic, left,
right and orientifold image of the
  right are respectively the dark solid, faint solid, dashed and dotted. The
  intersections corresponding to the quark doublets ($i = - 1, 0, 1$), up type
  singlets ($j = - 1, 0, 1$) and down type singlets ($j^{\ast} = - 1, 0, 1$)
  are denoted by an empty circle, full circle and a cross, respectively. All
  distance parameters are measured in units of $2 \pi R$ with $R$ the
  corresponding radius (except $\tilde{\epsilon}^{( 3 )}$ which is measured in
  units of $6 \pi R$).\label{figuratoros}}
\end{figure}

For the sake of clarity we will concentrate here on one very particular
model~\cite{Cremades:2003qj} that exemplifies most of the interesting
properties as well as some of the possible problems of models with branes
intersecting at angles. It is an orientifold compactification of type II
theory with four stacks of D6-branes wrapping factorizable 3-cycles on
the compact dimensions. This mouthfull is displayed in  Fig.~\ref{figuratoros}
which shows just the compactified space, $K_6$.
The compactified space is a compact factorizable 6-Torus
\[
T^2\times T^2 \times T^2,
\]
and the orientifold projection is given by $\Omega \mathcal{R}$ where $\Omega$
is the world-sheet parity and $\mathcal{R}$ is a reflection about the horizontal
axis of each of the three 2-tori,
\[
\mathcal{R} Z_I = \bar{Z}_I.
\]
We have denoted the coordinates of the tori by complex coordinates
$Z_I=X_{2I+2}+\mathrm{i} X_{2I+3}$, $I=1,2,3$, 
so the three boxes in the figure represent each 2 torus, 
with the edges being identified. 
Recall that the 6 branes must lie in $M_4$ so that there are only three
dimensions of each $D6$-brane that will appear in $K_6$. The branes therefore 
appear as just lines in each $T_2$. The net 
effect of the orientifold projection is to introduce mirror images of
the branes in each $T_2$  
(in the plane running horizontally). 

This particular model
contains at low energies just the particle content and
symmetries of the MSSM. In order to get that we include four stacks of
D6-branes, called \textit{baryonic} (a), \textit{left} (b), \textit{right}
(c), and \textit{leptonic} (d). Three of the dimensions of each D6-brane
wrap a 1-cycle on each of the three 2-tori, with wrapping numbers
denoted by $(n^I_k,m^I_k)$, \textit{i.e.} the stack $k$ wraps $n^I_k$ times
the horizontal dimension of the $I-$th torus and $m^I_k$ times the
vertical direction. We have to include for consistency their orientifold
images with $(n^I_k,-m^I_K)$ wrapping numbers. The number of
branes in each stack, their wrapping numbers and the gauge groups
they give rise to are shown in Table~\ref{data:branes:model} and a
subset of them, together with some of the relevant moduli, are displayed
in Fig.~\ref{figuratoros}.
\begin{table}[h]
\begin{center}
\begin{tabular}{|c|c|c|c|}
\hline
Stack &  $N_k$ & Gauge group & wrapping numbers
\\\hline
a & 3 & $\mathrm{SU}(3)\times \mathrm{U}(1)_a$ &(1,0);(1,3);(1,-3)
\\
b & 1 & $\mathrm{SU}(2)$ & (0,1);(1,0);(0,-1)
\\
c & 1 & $ \mathrm{U}(1)_c$ & (0,1);(0,-1);(1,0)
\\
d & 1 & $\mathrm{U}(1)_d$ &(1,0);(1,3);(1,-3)
\\\hline
\end{tabular}
\end{center}
\caption{
Number of branes, gauge groups and wrapping numbers for
the different stacks in the models discussed in the text.
\label{data:branes:model}}
\end{table}

The open string light spectrum in these models consists of the following fields:
\begin{itemize}
\item
$(p+1)$-dimensional gauge bosons (for the case of a stack of $N$ D$p$-branes)
corresponding in general to the group $\mathrm{U}(N)\sim \mathrm{SU}(N)
\times \mathrm{U}(1)$ live in the world volume of the corresponding
branes. In our particular configuration, we have seven-dimensional gauge
bosons corresponding to the gauge group $\mathrm{SU}(3)\times \mathrm{SU}(2)
\times \mathrm{U}(1)_a \times \mathrm{U}(1)_c \times \mathrm{U}(1)_d$ (see
Table~\ref{data:branes:model})~\footnote{Note that the left stack of branes
consists of just one brane that gives rise directly to a $\mathrm{USp}(2)
\sim \mathrm{SU}(2)$ gauge group instead of the usual $\mathrm{U}(1)$
due to the orientifold projection~\cite{Gimon:1996rq}.}.
Of the several abelian groups, every anomalous linear combination
receives a mass through the Green-Schwartz mechanism, whereas
anomaly-free combinations can remain massless or not, depending
on the particular brane configuration. This is indeed a salient feature of
this class of models that allow non-anomalous gauge bosons to couple to the RR
two-form fields acquiring a mass of the order of the string scale
in this form~\cite{Ibanez:2001nd}. The phenomenology of these
extra massive $\mathrm{U}(1)^\prime$s has been studied
in~\cite{Ghilencea:2002da} finding a bound on the string scale
$M_s \gtrsim 1$ TeV. Interestingly enough, these gauge symmetries
remain at the perturbative level as unbroken global symmetries~\cite{Ibanez:2001nd}.
Quite generally these new global symmetries correspond to baryon, lepton,
or Peccei-Quinn like symmetries, preventing proton decay even in low scale
models. In our particular example, the anomaly free massless combination
corresponding to the hypercharge is
\[
Q_Y=\frac{1}{6}Q_a-\frac{1}{2}(Q_c+Q_d).
\]

\item
Four-dimensional chiral massless fermions living on the intersections of
two branes and transforming as bi-fundamentals of the corresponding gauge
groups. Their number depend on a topological invariant, the intersection
number, which in the case of factorizable cycles on a factorizable
torus is simply
\[
\mathcal{I}_{ab}=\prod_{I=1}^3(n^I_am^I_b-m^I_a n^I_b),
\]
with different signs corresponding to different chiralities.
The fact that these branes wrap compact dimensions naturally
provide intersection numbers greater than one and therefore
replication of fermions with the same quantum numbers.
It should be mentioned here that in the case of lower-dimensional
branes, like D5 or D4-branes, chirality is not automatic and
locating the whole configuration at orbifold singularities is
required in order to get it~\cite{Aldazabal:2000dg}.

\item
Four-dimensional scalars, also localized at the intersections, with masses
that depend on the particular configurations of the branes. They can be seen
as the (generally massive when SUSY is broken by the intersection) superpartners
of the fermions at the intersections. In realistic models, scalars with the quantum
numbers of the (MS)SM Higgs boson also exist. In the example
we are considering the configuration is such that the same supersymmetry
is preserved at each of the intersections and massless scalars, superpartners
of the corresponding fermions completing the matter spectrum of the MSSM live
at the intersections.
\end{itemize}

The massive spectrum comprises, apart from the winding modes, that
correspond to stretched strings that wind around the compact
dimensions and have masses $M_{wind} \sim R_c/L_s^2$, where $R_c$ is
the compactification radius and $L_s$ the string scale,  KK
modes, that are states with non-zero (quantized in units of $1/R_c$
due to the periodicity conditions) momentum in the compact
dimensions and string excitations not related to the intersections
normally present in string models, a set of massive vector-like
fermions, the so-called 
\textit{gonions}~\cite{Aldazabal:2000dg}, localized near the intersections
and with angle-dependent masses. Although a purely effective field theory
study shows that relatively light vector-like fermions, especially when
they mix with the top quark, are the most likely source of modifications
of trilinear couplings~\cite{delAguila:2000aa}, the presence of Flavour
Changing Neutral Currents (FCNC) in these models overcomes in general any
phenomenological relevance of these states.

We have therefore seen that at the level of the light spectrum,
models with intersecting branes have a number of nice features, namely
four-dimensional chiral fermions, natural family replication and local
and global symmetries and matter content of the SM (or simple extensions 
thereof).
As we have seen, 
the closed string sector, which lives in the full ten-dimensional
target space, contains among other fields the graviton. These models
thus have a natural hierarchy of dimensionalities, with gravity propagating
in ten dimensions, gauge interactions in seven and matter in four. As we
sketched in the introduction, this will allow us to reduce the string
scale down to observable levels.

In our particular example, as can be seen in Fig.~\ref{figuratoros},
there are no dimensions transverse to \textit{all} the branes
and therefore no transverse volume can be made large enough to
account for the large effective four-dimensional Planck mass
with a small string scale. The thing that is stopping us are of course
the gauge couplings which would receive the same volume suppression
seen in Eq. (\ref{eq:ap}) and become extremely small. This problem can be circumvented
in several ways, the simplest one is to connect our small
torus to a large volume manifold without affecting the brane
structure~\cite{Blumenhagen:2002wn}, for instance, cutting
a hole and sewing and large volume manifold in a region away
from the branes~\footnote{There is a conceptual difficulty
in this construction that can be phrased as why in such a
large volume manifold, the relevant physics occurs in such
a tiny region. This difficulty is in one way or another always
present in the large extra dimensions approach to the hierarchy
problem but, as we have emphasized, the vacuum degeneracy problem
makes this possibility at least conceivable in a stringy set-up.}. 
This approach is in spirit quite similar to the bottom-up approach. 
A second possibility is to consider lower-dimensional branes, for
which transverse dimensions to all branes do exist. Realistic
examples with D5-branes and a string scale as low as
\textit{few} TeV have been constructed in~\cite{Cremades:2002dh}.
(See also~\cite{Bailin:2001ie} for other examples with
extra vector-like fermions.)
In these models the effective four-dimensional Planck mass
reads
\begin{equation}
M_{Pl}=\frac{2}{\lambda_{II}}M_s^4\sqrt{V_4 V_2},
\label{MPlanck:5D}
\end{equation}
where $V_{4,2}$ stand for the volume of the four-dimensional
manifold where the branes wrap and the volume of the
two-dimensional one transverse to all the branes and
$\lambda_{II}$ is the string coupling and $M_s$ is the
string scale. In this
situation it is possible to have all scales of order TeV but
the transverse dimensions then have to be $\sim$ \textit{mm}~\cite{add}.

Gauge couplings can be simply computed from a dimensional reduction of
the Yang-Mills theory living on the world-volume of the stack of
branes. As expected, it is suppressed by the volume of the compact
dimensions of the brane,
\begin{equation}
\frac{1}{g_a^2}=\frac{M_s^3}{16\pi^4 \lambda_{II}} V_a,
\label{gauge:couplings:D6}
\end{equation}
where we have considered the case at hand, \textit{i.e.} D6-branes
wrapping 3-cycles on the compact space and considered the gauge
coupling of an $\mathrm{SU}(N_a)$ group. Reasonable
values for the couplings are obtained if the relevant
volume for the brane is $V_a\sim M_{Pl}^3 \sim \mathrm{TeV}^3$.
Contrary to the original expectation, under certain mild assumptions,
gauge coupling unification can be obtained~\cite{Blumenhagen:2003jy}
(see also~\cite{Lust:2003ky} for a study of gauge threshold
corrections in intersecting brane models).

Models with intersecting branes therefore allow in principle
for a very low string scale, $M_s\sim 1$ TeV, while keeping
the Planck mass (\ref{MPlanck:5D}) and the gauge couplings
(\ref{gauge:couplings:D6}) at the observed values. Notice as
well that in the case of non-supersymmetric models, a low
string scale is preferred to avoid large corrections to the
Higgs vev.

We have not elaborated on the details of the construction and their consistency
conditions such 
as the absence of Ramond-Ramond tadpoles or the presence of unbroken
supersymmetries. These conditions greatly restrict the number of possibilities,
usually requiring the presence of more complicated spaces by further orbifolding
the toroidal structure we have discussed. See for instance~\cite{Uranga:2003pz}
for a nice review of this and other related topics.

Among the many phenomenological implications of low scale
models, flavour physics is one of the most pressing, so it is to flavour that we now turn. Flavour
experiments are typically able to probe mass scales much higher
than the energy of current experiments and as we will see shortly
this is particularly true in the case of intersecting brane models.
The flavour structure of these models is not restricted to Yukawa
couplings but flavour violating four-fermion contact interactions
are also present at the classical level, giving them a quite
unique rich structure. Nonetheless, since both sources of flavour
violation are intimately related we shall start with the description
of Yukawa couplings.

\subsection{Yukawa couplings}

\begin{figure}[h]
\begin{center}
\epsfig{file=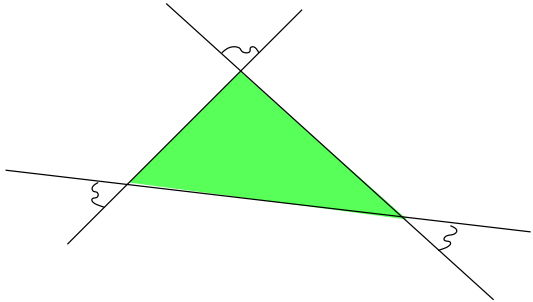,height=4cm}
\end{center}
\caption{World-sheet instanton contribution to the Yukawa couplings.
At each intersection a fermion or a scalar is localized. \label{3pt}}
\end{figure}

The leading contribution to Yukawa couplings between two fermions
and a scalar, each living at a different intersection, is due to
world-sheet instantons~\cite{Aldazabal:2000dg}. One can think of this as the 
classical action for a stretched string leaving an intersection 
(with one end on each brane) and travelling to the opposite corners of 
the Yukawa triangle. The action for a 
string is the worldsheet area, and therefore the amplitude should depend on the area the string sweeps out;
\begin{equation}
Y_{ijk}\sim \mathrm{e}^{-A_{ijk}/\alpha^\prime},
\end{equation}
where $A_{ijk}$ is the area of the minimal area worldsheet with vertices at
the three intersections, bounded by the corresponding branes.
(See Fig.~\ref{3pt}.)
A more detailed study of Yukawa couplings, using calibrated geometry
\cite{Cremades:2003qj}, and confirmed later by conformal field theory
techniques \cite{Cvetic:2003ch}, showed that when the compact space
is a factorizable torus and the branes wrap factorizable cycles,
the relevant area is the sum of the projected areas of the triangle
over each sub-torus. The final result, including the quantum part reads
\begin{equation}
Y=\sqrt{2} \lambda_{II} 2 \pi \sum_{I=1}^3 \sqrt{\frac{4\pi B(\nu_I,1-\nu_I)}
{B(\nu_I,\theta_I)F(\nu_I,1-\nu_I-\theta_I)}}\sum_m
\mathrm{e}^{-\frac{A_I(m)}{2\pi \alpha^\prime}},
\end{equation}
where we have neglected the presence of non-zero $B$ field and
Wilson lines and $B$ is the Euler Beta function, $I$ runs over the three tori,
$\nu_I$ and $\theta_I$ are the angles at the fermionic intersections,
$m$ runs over all possible triangles connecting the three vertices
on each of the three tori (there is an infinite number of them due
to the toroidal periodicity) and $A_I(m)$ is the projected area of the
$m-$th triangle on the $I-$th torus.

This exponential dependence has been claimed as a nice feature of these
models since it is expected to naturally give a hierarchical pattern
of fermion masses. As we shall see, in practice this does not hold, at least in the simplest models. The reason is that in many
cases, the dynamics of left-handed and right-handed fermions turns
out to occur in different tori and the property that only the projected
triangles are relevant translates into a factorization of the Yukawa
couplings. An example is the very model we have
been discussing in this section and displayed in detail 
in Fig.~\ref{figuratoros}.
Left-handed quarks live at different points only in the second torus
while they live at the same unique intersection in the third one.
The opposite happens for right-handed quarks. This results in the
following factorizable form of the Yukawa couplings
\begin{equation}
Y^u_{ij}= a_i b^u_j, \quad Y^d_{ij}=a_i b^d_j,
\label{factorisableyukawas}
\end{equation}
where we have only explicitly written the classical part, including this
time the presence of non-zero $B-$field and Wilson lines. The coefficients
are
\begin{eqnarray}
a_i&\equiv &  \vartheta \left[ \begin{array}{c} \frac{i}{3}+\epsilon^{(2)} \\ \theta^{(2)}
\end{array}\right] \Big( \frac{3 J^{(2)}}{\alpha^\prime} \Big),
\label{ai}
\\
b^u_j&\equiv &  \vartheta \left[ \begin{array}{c} \frac{j}{3}+\epsilon^{(3)}+\tilde{\epsilon}^{(3)} \\
\theta^{(3)} + \tilde{\theta}^{(3)}
\end{array}\right] \Big( \frac{3 J^{(3)}}{\alpha^\prime} \Big),
\label{bj}
\\
b^d_j&\equiv &  \vartheta \left[ \begin{array}{c} \frac{j^\ast}{3}+\epsilon^{(3)}-\tilde{\epsilon}^{(3)} \\
\theta^{(3)} - \tilde{\theta}^{(3)}
\end{array}\right] \Big( \frac{3 J^{(3)}}{\alpha^\prime} \Big),
\label{btildej}
\end{eqnarray}
where $i,j,j^\ast=-1,0,1$, $J^{(k)}$ denotes the complex K\"ahler
structure of the $k-$th torus, $\theta^{(2)},\theta^{(3)},
\tilde{\theta}^{(3)}$ parameterize the Wilson lines
and $\vartheta$ is the complex theta function with characteristics,
defined as
\begin{equation}
\vartheta\left[\begin{array}{c} \delta \\ \phi \end{array}\right]
(\kappa)=\sum_{l\epsilon Z} \exp [ \pi \mathrm{i} (\delta+l)^2
\kappa + 2 \pi \mathrm{i}(\delta+l)\phi ].
\end{equation}

This factorizable form of the Yukawa couplings, Eq. (\ref{factorisableyukawas}),
is too simple to lead to a realistic fermion spectrum. It is a rank one matrix
with one massive and two massless eigenvalues. There are of course
different ways out of this, either by using a more complicated
(non-factorizable) compact manifold or by looking for configurations
of branes in which the left and right dynamics occur at the same torus.
An example of the latter has been provided recently in
\cite{Chamoun:2003pf}, where a three Higgs model with democratic
rather than hierarchical Yukawas is studied.
There is however another feature of these very simple models that
makes the naive assertion above invalid when quantum corrections
are taken into account. This new feature is the presence of FCNCs 
that propagate through quantum loops
to the otherwise trivial structure of Yukawa couplings, providing them
with enough complexity to give rise to a realistic set of fermion
masses and mixing angles~\footnote{Although not necessary for the
generation of fermion masses, these FCNC also affect the model in
\cite{Chamoun:2003pf} as well, and therefore similar bounds on the
string scale apply.}.

\subsection{Flavour Changing Neutral Currents}

We have emphasized in this review that, after the 2nd string revolution,
string theory greatly influenced (and in turn received some degree of inspiration from) field theory investigations, particularly in the area of
models with extra dimensions. We shall see a salient example of the
complementarity between string and field theory in extra dimensions
in this section. Models with intersecting D-branes are a stringy
realization of the brane world idea, in which four-dimensional fermions live
in the boundaries of extra dimensions where gauge bosons are allowed
to propagate, these latter dimensions being a further restriction 
to a submanifold of the full
space-time where gravity lives~\cite{add,Rubakov:bb}. One well known property
of brane worlds in which the different fermions live in separate points of the extra
dimensions, the split fermion scenario~\cite{Arkani-Hamed:1999dc},
is the appearance of FCNCs that tightly
constraint the compactification scale $M_C \gtrsim 10^{2-3}$ TeV
in the case of flat extra dimensions~\cite{Delgado:1999sv}~\footnote{
The particular localization properties of KK modes in warped scenarios
make the bounds in that case milder~\cite{Huber:2003tu}).}. (See also
\cite{DelAguila:2001pu} for a model with light vector-like fermions,
relevant for phenomenology despite this very large compactification scale.)
The origin of these FCNC can be simply traced to the fact that
Kaluza-Klein modes of the multi-dimensional gauge bosons, having a
non-trivial profile in the extra dimensions, couple in a different
way to the fermions localized at the different positions. Family
non-universal gauge bosons then induce FCNC in the fermion mass
eigenstate basis~\cite{Langacker:2000ju}. Gauge boson KK generated
FCNC are therefore expected from a purely field theory viewpoint
in models with intersecting D-branes. A string calculation of the
tree level four-fermion amplitude, which can be performed~\cite{Cvetic:2003ch}
using an extension of the conformal field theory techniques 
developed for the heterotic orbifolds~\cite{Dixon:1986qv},
indeed reproduces the field theory expectation. In addition, though, it
reveals a new purely stringy source of flavour violation in these models
mediated by string instantons~\cite{Abel:2003fk}. These are simply 
worldsheets that directly connect four fermions of different generations
living at different intersections in the same way that the Yukawas 
connected the higgs to two fermions. Again the suppression goes roughly 
as the area, so that one would expect the FCNC effect from this source to 
increase as the compactification length and hence worldsheet area decrease.

The KK mediated flavour violating four fermion interactions
are of the form,
\begin{equation}
O_{LL}^{(\vec{n})}= \frac{(c_{LL}^{(\vec{n})})_{abcd}}{M_n^2}
(\bar{\psi}_{aL} \gamma^\mu \psi_{bL})(\bar{\psi}_{cL} \gamma^\mu \psi_{dL}),
\end{equation}
with the following dependence of the coefficient
\begin{equation}
(c_{LL}^{(\vec{n})})_{abcd}\sim \delta^{-M_{\vec{n}}^2/M_s^2}
\sum_{ij} (U^\dagger_L)_{ai} (U_L)_{ib} (U^\dagger_L)_{cj} (U_L)_{jd}
\cos\Big[\vec{M}_{\vec{n}} \cdot (\vec{y}^L_i - \vec{y}^L_j)\Big].
\end{equation}
$U_L$ are the corresponding unitary matrices rotating current
eigenstates into mass eigenstates and $\delta$ is an order
one (but always larger) number that depends on the specific brane
configurations and represents the string smoothing of the KK
contribution at high energies which is generally divergent in the 
field theory calculation of the same effect. 
(Essentially, the string smoothing 
arises because the branes have a finite 
width of order the string length, 
and are therefore unable to excite modes of a shorter wavelength than this.) We have only written the
Left-Left contribution, the case of Right-Right and Left-Right
contributions is a straight-forward generalization of that one.
Note that in order to have FCNC it is essential that current
and mass eigenstates are not aligned (so that the rotation
matrices are non-trivial) and the different generations
are localized at separate points of the extra dimension
($y_i-y_j\neq 0$). The exponential smoothing
provided by the string dynamics, which is crucial in the case
of more than one extra dimensions where the sums over KK modes
typically diverge, has to be introduced by hand in a field-theory
approach. This is another example of the complementarity between
string and field theory. String theory automatically cuts-off the
contribution of KK modes heavier than the string scale. Therefore
the larger the ratio $R_c/L_s$, the bigger the number of KK modes
that contribute and the larger the effect is.

\begin{figure}[h]
\begin{center}
\epsfig{file=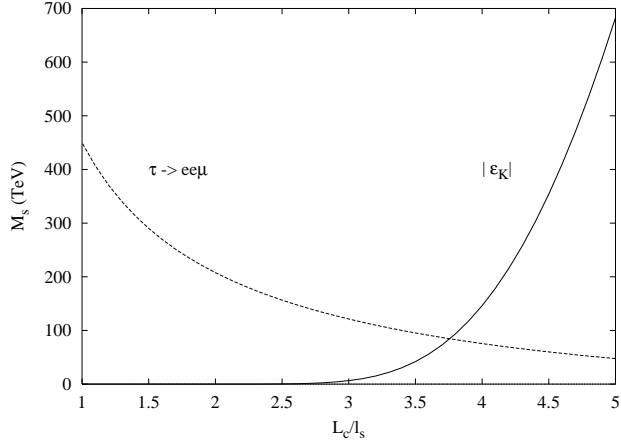,height=6cm}
\end{center}
\caption{Bound on the string scale as a function of the ratio $L_c/L_s$
from the KK contribution to $|\epsilon_K|$ and the string instanton
contribution to $\tau\to ee \mu$. A global bound $M_s\gtrsim 100$ TeV
is found. \label{globalbound:plot}}
\end{figure}

On the other hand, string instanton FCNCs
depend very much on the chiralities of the external
fermions (through the difference in the number of independent
angles). Four-fermion interactions with all fermions of the
same chirality (either all LH or all RH) correspond to a
parallelogram with only one independent angle. Given the
factorization property of the model we are discussing, the
only non-vanishing world-sheet area occurs in one torus
and the result is of the form
\begin{equation}
O_{LL}^{str}= \frac{(c_{LL}^{(\vec{n})})_{abcd}}{M_s^2}
(\bar{\psi}_{aL} \gamma^\mu \psi_{bL})(\bar{\psi}_{cL} \gamma^\mu \psi_{dL}),
\end{equation}
with the following dependence of the coefficient
\begin{equation}
(c_{LL}^{str})_{abcd}\sim
\mathrm{e}^{-\frac{A}{2\pi L_s^2}}
\sum_{i} (U^\dagger_L)_{ai} (U_L)_{(i+1)b} (U^\dagger_L)_{c(i+1)} (U_L)_{(i+2)d},
\end{equation}
where $A$ is the area of the corresponding parallelogram
(which is $\sim (4\pi^2 R_c^2)/3$) and $L_s=1/M_s$ is the string scale.
Already in this chirality preserving interaction we observe
several differences with respect to the field theory case.
The first one is that there are FCNC even in the case of
Yukawa couplings aligned with gauge couplings (\textit{i.e.}
$U=1$). Secondly, the exponential dependence on the ratio
of string and compactification scales is opposite to that coming from 
the KK modes,
the larger the ratio $R_c/L_S$, (\textit{i.e.} the larger
the area in string units) the stronger the suppression.
Notice however that it is still necessary to have different
generations living at separate points in order to have FCNC.
The opposite dependence of the KK and string instanton contributions
on the ratio of compactification and string scales allows us
to put a lower bound on the string scale, independently of
this ratio. An estimation of this bound~\cite{Abel:2003fk},
using the KK contribution
to $|\epsilon_K|$ and the string instanton contribution to
$\tau \to ee\mu$ and relatively small mixing angles, leads to
the bound $M_s\gtrsim 100$ TeV as shown in Fig.~\ref{globalbound:plot}.

The chirality changing four-fermion interactions, connecting
two left-handed and two right-handed fermions, is a bit
more involved but far more interesting. We will give the
final expressions here and outline the reasons for the
new features without entering into the intricacies of the
calculation. The main new feature is the absence of L-R
factorization in the amplitude (except in some limiting cases).
The reason is that now in general there are non-zero contributions
in more than one 2-torus and the classical
action is no longer the sum of the areas of each
of the quadrangles (incidentally, this does not happen for
the Yukawa couplings because in the three-point amplitude
we can fix all three vertices using $SL(2,R)$ invariance whereas
in the four-point one we have to integrate over the position
of the fourth vertex, see~\cite{Owen:SP03}.) As we shall see
soon, this introduces enough flavour violation  to generate,
through loop corrections, a semi-realistic pattern of fermion
masses and mixing angles.

\begin{figure}[h]
\begin{center}
\epsfig{file=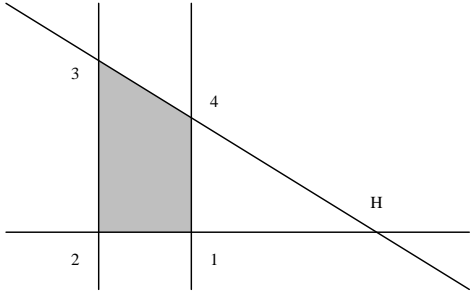,height=4cm}
\end{center}
\caption{Higgs vs string instanton mediation of the process
$(\bar{q}_{aL} q_{bR})(\bar{q}_{cR} q_{dL})$\label{Yuk:enhancement:fig}}
\end{figure}

Another nice feature with possible important phenomenological
implications is related to Higgs-mediated like processes.
Let us consider the situation displayed in Fig.~\ref{Yuk:enhancement:fig}.
The Higgs mediated process can be obtained as the field theory
limit of a string propagating from the vertices 2 and 3 down
to the Higgs vertex and then back to the vertices 1 and 4.
This contribution goes, in the $t$ channel, like
\begin{equation}
\frac{\mathrm{e}^{-A_{23H}/2\pi L_s^2}\mathrm{e}^{-A_{14H}/2\pi L_s^2}}
{t-M_H^2}\sim \frac{Y_{23} Y_{14}} {t-M_H^2},
\end{equation}
where $M_H$ is the Higgs mass. On the other hand there is
another, purely stringy contribution (not expected
on field theory grounds) that can be very much enhanced for
a low string scale and corresponds to a string sweeping out
the area of the quadrangle between the four vertices 1,2,3,4
without going through the Higgs vertex (shaded area in the Figure).
In this case if all the flavour dynamics happens on a single torus 
the amplitude goes as
\begin{equation}
\frac{\mathrm{e}^{-A_{1234}/2\pi L_s^2}}
{M_s^2}
\sim \frac{Y_{23}/Y_{14}} {M_s^2}.
\end{equation}
If the flavour dynamics happens in more than one torus  
the detailed result depends
on the particular configuration due to the non-factorization property
of this four point amplitude alluded to above, but is 
roughly the same.
A more detailed study is necessary before making any statement
about the phenomenological implications of this property but
it seems that a general feature of models with intersecting
branes is the presence of Higgs-like processes enhanced (as
opposite to the usual expected suppression) by light Yukawas.

Let us now concentrate on the relevant amplitude for the generation
of fermion masses and mixing angles. In particular we will consider the
quark sector and are interested on the $(\bar{q}_{aL} q_{bR})
(\bar{q}_{cR} q_{dL})$ amplitude. The full expressions are intricate
and do not admit a simple analytical form. In order to give some
feeling of what happens we will consider a simplified case
in which the relevant angles are the same on
each sub-torus. In this case the classical action turns out to be 
\cite{Cvetic:2003ch}
\begin{equation}
S_{cl}=\frac{1}{4\pi \alpha^\prime}
\frac{\sin \pi \vartheta_2\sin \pi \vartheta_3}{\sin (\pi \vartheta_2
+\pi \vartheta_3)}
\sqrt{\sum_m(v_{23}^m-v_{14}^m)^2\sum_n(v_{23}^n-v_{14}^n)^2},
\label{Scl:LR}
\end{equation}
where $\theta_{2,3}$ are the (independent) angles at the corresponding intersections
and $v_{23}$, $v_{14}$ are the distances between the relevant intersections.
From this expression it is clear that only in the trivial case
(when $a=d$ or $b=c$) or in the degenerate case (when distances
in all sub-tori are equal) the amplitude $\sim \exp(-S_{cl})$
factorizes.

We have discussed the different tree-level contributions to
the flavour structure of models with intersecting branes.
Let us consider now how this highly non-trivial structure
of flavour violation propagates, through quantum corrections
to the otherwise trivial (at tree level) Yukawa couplings.

\subsection{Yukawa couplings at one loop}

\begin{figure}[h]
\begin{center}
\epsfig{file=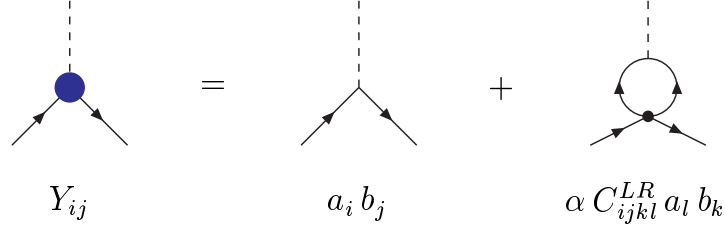,height=3cm}
\end{center}
\caption{One loop threshold correction to Yukawa couplings.
The black dot represents the insertion of a flavour violating
four-fermion contact interaction.\label{oneloopYuk}}
\end{figure}

Let us recapitulate the main features of the model we are
considering. Tree level Yukawa couplings factorize in left
and right parts, leading to a rank one matrix of the form
\begin{equation}
Y^{u}_{ij}=a_i b^u_j, \quad
Y^{d}_{ij}=a_i b^d_j.
\end{equation}
On the other hand, four-fermion contact interactions
violate flavour at tree level. In particular there is a
chirality changing contribution that does not factorize
except at particularly symmetric points. One then expects
that this non-trivial flavour structure propagates at
the quantum level to the Yukawa couplings as sketched
in Fig.~\ref{oneloopYuk}. The expected one loop value
or the Yukawa couplings is
\begin{equation}
Y_{ij} \sim a_i b_j + \alpha C^{LR}_{ijkl} a_l b_k + \ldots
\end{equation}
where we have loosely denoted by $\alpha$ the loop suppression
and $C^{LR}_{ijkl}\sim \exp(-S_{cl})$ with the classical action
similar to the one in Eq. (\ref{Scl:LR}).

Note that there is another one loop contribution to the Yukawa
couplings mediated by KK gauge bosons that we are neglecting for
simplicity~\footnote{It is indeed expected to give a small
correction given the fact that, in order to have a large enough
top Yukawa, the string scale has to be close to the compactification
scale and therefore the string suppression of KK couplings is
very effective.}. This and other, higher loop, corrections
would be necessary in the case that the LR string contribution
factorized, the Yukawa coupling would be in this case still
a rank two matrix and therefore those corrections would be
essential to give masses to the first generation. In general
this contribution does not factorize though and we can generate
the full fermionic spectrum with just this leading effect.
In order to get approximate analytical expressions we will
assume that there is a small deviation on $C^{LR}$ from
factorization,
\begin{equation}
C^{LR}_{ijkl} a_l b_k= c_i d_j + \epsilon \tilde{C}^{LR}_{ij},
\end{equation}
where $\tilde{C}^{LR}$ is a general matrix and $\epsilon$ is
a small parameter.
The full Yukawa matrix can then be written, at one loop order,
as
\begin{equation}
Y^{u,d}_{ij}=a_i b^{u,d}_j+ \alpha c_i d^{u,d}_j +
\alpha \epsilon \tilde{C}^{u,dLR}_{ij}.
\end{equation}
This matrix can be perturbatively diagonalized by
the following unitary matrices,
\begin{equation}
L^{u,d}=L_0
\begin{pmatrix}
1-\Big(\frac{\mu_{12}^{u,d}}{\mu_1^{u,d}}\Big)^2 \epsilon^2 &
\frac{\mu_{12}^{u,d}}{\mu_1^{u,d}} \epsilon &
\alpha \epsilon \frac{\mu^{u,d}_{13}}{|a||b^{u,d}|} \\
-\frac{\mu_{12}^{u,d}}{\mu_1^{u,d}} \epsilon &
1-\Big(\frac{\mu_{12}^{u,d}}{\mu_1^{u,d}}\Big)^2 \epsilon^2 &
\alpha \frac{\mu^{u,d}_2}{|a||b^{u,d}|} \\
\alpha \epsilon \frac{\mu^{u,d}_{12}\mu^{u,d}_{2}/\mu^{u,d}_{1}-\mu^{u,d}_{13}}
{|a||b^{u,d}|}&
-\alpha \frac{\mu^{u,d}_2}{|a||b^{u,d}|} &
1
\end{pmatrix},
\label{Lrot}
\end{equation}
and a similar rotation for the right handed fields. The order
one rotation reads
\begin{equation}
L_0=\Big( \frac{\vec{a}\wedge\vec{c}}{|\vec{a}\wedge\vec{c}|},
\frac{(\vec{a}\cdot \vec{c}) \vec{a}-a^2 \vec{c}}{|(\vec{a}\cdot \vec{c}) \vec{a}-a^2 \vec{c}|},
\frac{\vec{a}}{|a|}\Big),
\end{equation}
and the resulting mass eigenstates are, to leading order,
\begin{equation}
M^{u,d}=\frac{v^{u,d}}{\sqrt{2}}\begin{pmatrix}
\alpha \epsilon \mu^{u,d}_{11} && \\
&\alpha \mu^{u,d}_1 & \\
&& |a||b^{u,d}|
\end{pmatrix},
\end{equation}
where we have included the vev of the up or down Higgs, related
by $\tan \beta$.
Finally, the $(12), (13)$ and $(23)$ entries of the CKM miximg matrix,
$V^{CKM}=L^{u\,\dagger}L^d$, are
\begin{align}
V^{CKM}_{12}=&
\epsilon \Big[ \frac{\mu^d_{12}}{\mu^d_1}-\frac{\mu^u_{12}}{\mu^u_1}\Big],
\nonumber \\
V^{CKM}_{13}=&
\alpha \epsilon \frac{1}{|a|}
\Big[
\frac{\mu^{u}_{12}\mu^{u}_{2}/\mu^{u}_{1}-\mu^{u}_{13}}
{|b^{u}|}
-
\frac{\mu^{u}_{12}\mu^{d}_{2}/\mu^{u}_{1}-\mu^{d}_{13}}
{|b^{d}|}
\Big]
\nonumber \\
V^{CKM}_{23}=&
\alpha \frac{1}{|a|} \Big[
\frac{\mu^d_2}{|b^d|}-\frac{\mu^u_2}{|b^u|}
\Big].
\nonumber
\end{align}
The different coefficients in the previous equations are
\begin{equation}
\mu_1^{u,d}= \frac{|\vec{a}\wedge \vec{c}|
|\vec{b}^{u,d}\wedge \vec{d}^{u,d}|}{|\vec{a}||\vec{b}^{u,d}|},
\quad\mu_2^{u,d}=- \frac{|\vec{a}\wedge \vec{c}|
(\vec{b}^{u,d}\cdot \vec{d}^{u,d})}{|\vec{a}||\vec{b}^{u,d}|},
\end{equation}
and
\begin{equation}
\mu^{u,d}_{ij}=(L_0^\dagger \tilde{C}^{LR} R_0)_{ij},
\end{equation}
all expected to be order one.
The hierarchical pattern of quark masses and mixing angles found
in nature~\cite{PDG}
\begin{align}
&m_u\sim 3\times 10^{-3} \mbox{ GeV},\quad  m_c\sim 1.2\mbox{ Gev},
\quad  m_t\sim 174 \mbox{ GeV}, \nonumber \\
&m_d\sim 7\times 10^{-3} \mbox{ GeV},\quad m_s\sim 0.12\mbox{ Gev},
\quad  m_b\sim 4.2 \mbox{ GeV},\label{massesandmixings:data}\\
& V_{12}\sim 0.22, \quad V_{13}\sim 0.0035, \quad V_{23}\sim 0.04,
\nonumber
\end{align}
can be explained by a hierarchy in the expansion coefficients,
$\alpha$ and $\epsilon$. In fact reasonable values for all experimental
data in Eq. (\ref{massesandmixings:data}), up to order one coefficients,
can be obtained using,
\begin{equation}
\alpha\sim 10^{-2}, \quad \epsilon \sim 0.1,
\end{equation}
but for the up quark for which some amount of cancelation seems
necessary.

\subsection{Experimental bounds on the string scale}

Once we have developed a semi-realistic theory of flavour
in a concrete model with intersecting branes we can estimate
the contribution to flavour violating processes (such as rare
decays, meson oscillations, etc.) and extract from them
stringent experimental bounds on the
string scale for these models. Although a definite pattern
for the fermion spectrum along the lines outline above
has not yet been fully developed, estimates using the
one and two loops KK contribution to the Yukawa couplings
\cite{Santiago:SP03}, leads to the bounds on the string scale
shown in Table~\ref{Msbounds}.

\begin{table}[h]
\begin{center}
\begin{tabular}{||c|c|||c|c||}
\multicolumn{2}{c}{Quark sector}&
\multicolumn{2}{c}{Semileptonic Observables}\\
\hline Observable & $M_s \gtrsim$ (TeV)&
Observable & $M_s \gtrsim$ (TeV)\\
\hline
$\Delta m_K$ & 4000 &
$\mu-e$ conversion & 2700\\
\hline
$\Delta m_{B_d}$ & 1300 &
$K\to \mu \mu $ & 80 \\
\hline
$\Delta m_{B_s}$ &  500
&
$K \to \pi \mu \mu $ & 200
\\
\hline
$\Delta m_D$ & 2000&
Supernovi & 10
\\ \hline \hline
\small $|\epsilon_K|$ & $10^4$ &&\\
\hline
Hg EDM & 10 &&
\\\hline
\end{tabular}
\end{center}
\caption{Bounds on the string scale from different observables
in the quark (left) and leptonic (right) sectors. The observables
in the upper part are CP preserving while the ones in the lower
part of the table are CP violating.
 \label{Msbounds}}
\end{table}

In this table CP conserving quark observables are considered in the
upper left side. In the lower left side, we include quark CP violating
observables whereas the right side is devoted to semileptonic
observables. The bounds should be taken with caution. First,
a fully realistic example of fermion masses and mixing angles
generation along the lines above has not been produced yet and
the detailed value of the FCNC present in the model depends as
we saw on the rotation matrices and therefore on the Yukawa
couplings~\footnote{In Ref.~\cite{Chamoun:2003pf} a detailed
account of the fermionic spectrum is provided. The authors however
consider an intermediate string scale and therefore do not bother
about FCNC which would be anyway irrelevant in their case.}.
Second, in the estimates of Ref.~\cite{Santiago:SP03}, only
the quark sector was considered, and CP violation was neglected.
This means that the very stringent bounds should be taken as
estimates of the order of magnitude of the result in a more
realistic calculation, and are more precise in the CP conserving
quark sector than in the rest. Nevertheless it is clear
that the bounds are so constraining that it does not seem
feasible to have models of intersecting branes with a very
low string scale. Equivalently we can say that flavour observables
are probing string scales of the order of $M_s \gtrsim 10^{3-4}$ TeV.

These bounds have been obtained for a very particular intersecting
brane model. The presence of FCNC is however quite general in these
models (unless we require all three families to live at the same
intersection in which case we loose some of the nice features of these
models such as family replication of generations or hierarchical Yukawa
couplings) and although model dependent,  it is natural to expect a bound on the string scale of
the same order of magnitude for a wide variety of models with branes
intersecting at non-trivial angles, because of 
the different origin and
moduli dependence of the various sources of flavour violation together
with the fact that flavour-violating observables are so restricted
experimentally.
Therefore we seem to be forced back to high string scales solely on
experimental grounds. A great deal of effort has been dedicated
to the study of realistic supersymmetric models (in order to
protect the hierarchy of scales) with a high string scale, of the
order of the Planck or GUT scales~\cite{Cvetic:2001tj,Kokorelis:2002ip} and 
it seems likely that it is to these that we must now turn.

\section{Conclusions}

In this paper we have described in some detail the way in which string
theory has been able to accommodate the most esoteric of theoretical
ideas, that of large extra dimensions or low fundamental scales. Here
the approach we have taken is to show how string theory has developed
alongside the more generic ideas to do with extra dimensions that
have been proposed in a purely field theoretic set-up. Beginning with
the Horava-Witten set-up which incorporated a fundamental scale of
$\sim 10^{16}$ GeV, the construction of models involving D-branes has
allowed set-ups first with an intermediate fundamental scale of $\sim 10^{11}$ GeV
and later with fundamental scales of $\sim 1$ TeV. The latter realized
in string theory the ideas put forward in ref. \cite{add}
for solving the hierarchy problem. The large Planck scale was in this
scenario put down to the large volume of some compactified space rather
than any fundamental hierarchies of scale. The question of fine-tuning
could then be shifted onto the various moduli fields that describe
how large the compact space is, where (it is hoped) one might be able
to exercise more control.

There are a number of aspects that we have touched on and that we
would like to reemphasize here. The most important we feel is the
fact that the stringy constructions have had to conform to the strictures
imposed upon them by the various dualities inherent in string theory.
This makes them far more constrained than might have been expected
and certainly more constrained than the field theory equivalents.
For example we have shown how flavour changing (FCNC) experiments
in fact rule out the low scale string models and actually probe
models with string scales all the way up to $10^{7}$ GeV. In a sense
this makes the string approach more honest. In a field theoretic set-up
it often seems to be possible to avoid experimental bounds by retiring
to some corner of parameter space. In string theory such corners tend
not to exist. A good example of this is the FCNC effect of Kaluza-Klein
modes that were considered. In field theory one can reduce the size
of the relevant dimensions to make the modes heavy and turn this source
of FCNC off. In string theory however this merely introduces compensating
FCNCs due to string instanton effects.

One further aspect we touched upon in the context of the intermediate
models, is that stringy set-ups may introduce a reasonable explanation
for some parameters that are apparently fine-tuned (i.e. unnatural
in the sense of t'Hooft). The particular example we chose was the
phenomenon of Kinetic-Mixing leading to just the right mediation of
D-term supersymmetry breaking. Although this is encouraging, we think
that one has to be rather careful in the interpretation; there is
no such thing as a free lunch. What really happened in this case is
that the fine-tuning of the Kinetic-Mixing was tied to the volume
suppression of the gravitational interactions. In the end the number
of fine-tunings is reduced but one is still left with the problem
of fine-tuning the large volume. This aspect of large extra-dimensions
and low fundamental scales has to wait until a better understanding
of the behaviour of moduli fields and in particular supersymmetry
breaking.

\section*{Acknowledgements}
It is a pleasure to thank Oleg Lebedev, 
Manel Masip, Anthony Owen 
and Ben Schofield for their comments, discussion and collaboration.
This work has been partially supported by PPARC Opportunity Grant 
PPA/T/S/1998/00833.

\end{document}